# Demographics in Social Media Data for Public Health Research: Does it matter?

Nina Cesare
Institute for Health Metrics and Evaluation
University of Washington
Seattle, Washington, USA
ninac2@uw.edu

Christan Grant
School of Computer Science
University of Oklahoma
Norman, OK, USA
cgrant@ou.edu

Jared B. Hawkins
Boston Children's Hospital
Harvard Medical School
Boston, MA, USA
jared.hawkins@childrens.harvard.edu

John S. Brownstein
Boston Children's Hospital
Harvard Medical School
Boston, MA, USA
john.brownstein@childrens.harvard.edu

Elaine O. Nsoesie
Institute for Health Metrics and Evaluation
University of Washington
Seattle, WA, USA
en22@uw.edu

## ABSTRACT

Social media data provides propitious opportunities for public health research. However, studies suggest that disparities may exist in the representation of certain populations (e.g., people of lower socioeconomic status). To quantify and address these disparities in population representation, we need demographic information, which is usually missing from most social media platforms. Here, we propose an ensemble approach for inferring demographics from social media data.

Several methods have been proposed for inferring demographic attributes such as, age, gender and race/ethnicity. However, most of these methods require large volumes of data, which makes their application to large scale studies challenging. We develop a scalable approach that relies only on user names to predict gender. We develop three separate classifiers trained on data containing the gender labels of 7,953 Twitter users from Kaggle.com. Next, we combine predictions from the individual classifiers using a stacked generalization technique and apply the ensemble classifier to a dataset of 36,085 geotagged foodborne illness related tweets from the United States.

Our ensemble approach achieves an accuracy, precision, recall, and F1 score of 0.828, 0.851, 0.852 and 0.837, respectively, higher than the individual machine learning approaches. The ensemble classifier also covers any user with an alphanumeric name, while the data matching approach, which achieves an accuracy of 0.917, only covers 67% of users. Application of our method to reports of foodborne illness in the United States highlights disparities in tweeting by gender and shows that counties with a high volume of foodborne-illness related tweets are heavily overrepresented by female Twitter users.

Traditional public health data typically includes demographic information, such as age, gender and race. While social media can supplement traditional public health surveillance systems, its limitations (e.g., absence of user demographics) have not been properly quantified and addressed via statistical methods or other approaches. We argue that understanding disparities in this data can aid in addressing the underrepresentation of certain populations. Ongoing work is focused on extending the ensemble method and applying a similar approach to predicting age and race.

## 1. INTRODUCTION

Public health researchers have used data from social media to characterize attitudes towards vaccines [5, 47], tobacco use [18, 25, 35, 44], and quality of care in hospitals [23]. These data have also been used to track disease outbreaks and reports of illness [7, 14, 15, 21, 40, 46, 48], to study mental health [51], [17], to analyze sleep habits [31], to assess neighborhood trends in diet and weight loss [37, 42], to measure the geographic distribution of fitness activity [22] and to map the presence of food deserts [16]. Some of the advantages of using these data for public health research include timeliness, and affordability [14, 43, 49].

Despite these advantages, research suggests that there may be disparities in populations that use digital tools for sharing personal health data. For instance, Henly et al. [24] analyzed the relationship between reporting foodborne illness on Yelp and socioeconomic status (SES) and found that counties with lower SES were less likely to report foodborne illness. Other studies have documented demographic bias in the use of digital health apps [8, 11, 20], as well as disparities in seeking health information through digital sources [29]. Potential reasons for these disparities include, lack of interest in or time for sharing personal health information online, limited access to technology, and distrust.

To quantify and address demographic disparities in social media data, demographic data is needed. We argue that estimating social media users' demographics would enable researchers to assess the quality of social media data and to develop statistical methods that adjust for data bias. Since these data are being used for public health research and applications, understanding disparities in population representation would ensure that existing health disparities are not replicated or magnified through online environments.

However, most social media platforms do not provide demographic details for users. This has motivated the proposal of methods for predicting the demographics of social media users. Proposed approaches range from matching components of the user's profile (e.g., the user's name) to public data sources (e.g., the U.S. Census), to supervised learning approaches that use profile content and/or text posted by the user to predict demographic features. However, most of these methods require large volumes of data, which makes the application to large scale studies challenging. Here, we propose an ensemble approach that combines three simple methods for inferring the gender of Twitter users using only their listed names.

In the remainder of this paper we present, (1) a review of approaches for estimating user demographics, (2) describe three scalable approaches for predicting Twitter users' gender, (3) introduce a weighted ensemble approach for combining these individual methods, and (4) apply these methods to tweets about foodborne-related illness to examine gender disparities in reporting in the United States at the county level. We conclude by discussing the implications of these findings for public health research.

## 2. METHODS
### 2.1. APPROACHES FOR DEMOGRAPHIC DETECTION

We previously performed a comprehensive review of methods for inferring demographics of social media users (see Cesare et al. [12]). We identified 60 studies; 47 predicted gender, 29 predicted age, and 13 predicted race or ethnicity. The proposed approaches used a variety of techniques, including human and automated facial recognition, simple and adjusted data matching, Bayesian estimation, and unsupervised and supervised learning.

The distribution of papers reviewed, sorted by year of publication and platform analyzed is displayed in Figure 1. Data from Twitter was most frequently studied. Also, most studies prioritized increasing prediction accuracy, which usually involved the collection and analysis of detailed user metadata, including text from users' posts and network ties. Approaches reliant on posted text can be computationally demanding and not scalable [1, 2]. Additionally, including text and/or network data does not always significantly improve classification [6]. Researchers working on frequently discussed health topics – such as exercise and wellness across U.S. counties [50] – may find it difficult to process and analyze texts posts and network ties for every user in their sample. Also, researchers studying time-sensitive phenomena – such as an infectious disease outbreak [13, 14] – may not have the time necessary to collect text posts or network ties for their entire sample. Further discussions on scalability and efficiency can be found in the review manuscript, Cesare et al. [12].

### 2.2. GENDER DETECTION METHODS
We propose a scalable, weighted ensemble classification framework for predicting Twitter users' gender using only their names. We considered five approaches: a data matching approach, an approach using facial recognition technology, and three supervised learning approaches. We implemented each method and incorporated predictions from the top performing approaches into an ensemble classifier using a weighted ensemble classifier.

To train and test each classification method, we used Twitter data with gender labels generated by the crowdsourcing platform Crowdflower[1] and made available through Kaggle[2] - a platform that shares data from companies and researchers and invites statisticians and data scientists to use the data for predictive modeling. The data used for this project included 20,000 tweets selected at random via Twitter's streaming API. Three Crowdflower workers evaluated each user profile to determine their gender. Because Crowdflower algorithmically generated a confidence score for these evaluations,[3] we retained only users whose gender was estimated with one hundred percent confidence. We also eliminated any users not coded as male or female (e.g. "brand" or "unknown"). We then extracted full user metadata for each of these users through Twitter's REST API. A total of 7,953 users were included in our gender prediction training and test data.

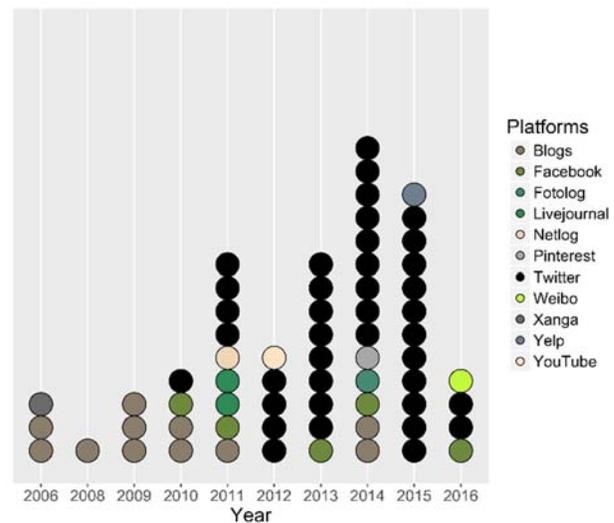

**Figure 1: Number of research articles focused on each platform by year**

### 2.2.1. METHOD 1: MATCHING NAMES TO US SOCIAL SECURITY ADMINISTRATION DATA
Our first approach to gender prediction matched users' first names to historical data from the US Social Security Administration (SSA). Mislove et al. [32] found that approximately 67% of a random sample of US Twitter users elected to provide a real first name in their profile. However, first names are often embedded in other text and must be extracted and cleaned prior to analysis. We adopted an approach similar to that used by Longley et al. [27] to extract first names from user names. The process involved: (1) removing suffixes (e.g. Mrs, Dr, etc.) and stop words ("and," "the," etc.), (2) converting the names from ASCII to UTF-8 to remove unusual characters, (3) normalizing all characters, (4) trimming trailing and leading white space, (5) removing numbers and (6) removing punctuation, underscores, and Unicode characters. If the result was a unigram, it was retained. If the result was a bigram or longer, we kept the string that preceded the last string, which we assumed to be the surname (i.e. "Lil' John *Doe*").

---
[1] https://www.crowdflower.com/
[2] Data are available here: https://www.kaggle.com/crowdflower/twitter-user-gender-classification
[3] https://success.crowdflower.com/hc/en-us/articles/201855939-How-to-Calculate-a-Confidence-Score

The resulting name unigrams were then processed using the *gender* package in the R statistical program[4]. This package searches input names within a historical database of SSA data (from 1940 onward), and estimates the probability that the names belong to male or female users. Table 1 displays the performance measures for this gender detection technique.

### 2.2.2. METHOD 2: SUPERVISED LEARNING USING WORD AND CHARACTER N-GRAMS

This method is motivated by Burger et al. [10], who used word and character n-grams from profile text fields and tweets to predict gender. Similar to Burger et al. [10], we initially tested algorithms that used n-grams from a variety of profile text fields - including description, name, screen name and last status posted – to predict gender. We consistently found that models using word and character n-grams from user names provided the best predictions. Table 1 summarizes the performance of this approach across five algorithms. While logistic regression performed well, we encountered issues with collinearity when bootstrap resampling due to the large number of features used. We therefore chose to use Support Vector Machines (SVM) since it achieved a similar level of performance.

### 2.2.3. METHOD 3: SUPERVISED LEARNING AND LINGUISTIC NAME STRUCTURE

The third method used to detect gender replicated an approach proposed by Mueller and Strumme [33]. These authors highlighted that while first names are useful for predicting gender, names on Twitter are often ill-formed and difficult to identify (for instance, "McKayla" may jokingly be spelled "McKaalyaaaa" or "Mah-Kaylaaaa"). Even with thorough cleaning and pre-processing, the gender of some names might not be detectable through a data matching approach. However, they suggested that even with linguistic variation, male and female names may have different structures that enable the detection of a user's gender.

We first extracted the user's first name as described in method 1. Next, we extracted the following predictive features: the number of syllables in the name, the number of vowels and consonants in the name, the number of bouba and kiki vowels and consonants in the name [30, 38], and a binary measure of whether the last character in the name is a vowel. We compared the performance of five machine learning algorithms in predicting gender using these features (see Table 1).

### 2.2.4. WEIGHTED ENSEMBLE GENDER CLASSIFIER

We chose a weighted ensemble classification scheme that combines the output of the three methods highlighted using a stacked generalization approach [54]. Ensemble classifiers tend to outperform individual classifiers when working with noisy data [39, 52], and user-generated Twitter data features significant variation.

First, we used identical test datasets for evaluating each algorithm (with methods 2 and 3 calibrated on the same training set). We then fit a logistic regression model whose inputs are the gender estimates of each algorithm given the test data, and whose output is the set of ground truth gender measures from the test data. Due to the limited coverage of Method 1, we developed two weighted ensemble models – one that included users whose gender was predicted by Methods 2 and 3, and another that included users whose gender was predicted by all three methods. The structure of this approach is illustrated in Figure 3. This approach achieves an accuracy of 0.828, precision of 0.851, recall of 0.852, and F1 score of 0.837. Most importantly, it can be applied to nearly all users in the sample.

To illustrate the scalability of our approach and a potential application, we applied the ensemble classifier to a dataset of 36,085 geotagged tweets of symptoms of foodborne illness posted by 33,208 users in the US during 2013. We analyzed the total volume of tweets originating from each US county, and the proportion of tweets posted specifically by female users.

**Table 1: Performance measures for each prediction method**

|  | Algorithm | Accuracy | Precision | Recall |
|---|---|---|---|---|
| Method 1: SSA data | Data matching | 0.917 | 0.941 | 0.893 |
| Method 2: N-grams | SVM (linear kernel) | 0.752 | 0.777 | 0.671 |
|  | Naïve Bayes | 0.472 | 0.476 | 0.992 |
|  | Decision tree | 0.626 | 0.614 | 0.606 |
|  | Balanced Winnow | 0.669 | 0.652 | 0.674 |
|  | Logistic regression | 0.756 | 0.759 | 0.725 |
| Method 3: Linguistic structure | SVM (radial kernel) | 0.644 | 0.626 | 0.711 |
|  | Naïve Bayes | 0.583 | 0.564 | 0.717 |
|  | Decision tree | 0.698 | 0.695 | 0.701 |
|  | Balanced Winnow | 0.690 | 0.681 | 0.712 |
|  | Logistic regression | 0.635 | 0.625 | 0.67 |
|  |  | F1 | Coverage |  |
| Method 1: SSA data | Data matching | 0.916 | 68% |  |
| Method 2: N-grams | SVM | 0.726 | 100% |  |
|  | Naïve Bayes | 0.628 |  |  |
|  | Decision tree | 0.610 |  |  |
|  | Balanced Winnow | 0.663 |  |  |
|  | Logistic regression | 0.742 |  |  |
| Method 3: Linguistic structure | SVM | 0.666 | 100% |  |
|  | Naïve Bayes | 0.632 |  |  |
|  | Decision tree | 0.698 |  |  |
|  | Balanced Winnow | 0.696 |  |  |
|  | Logistic regression | 0.647 |  |  |

---

[4] https://cran.r-project.org/web/packages/gender/

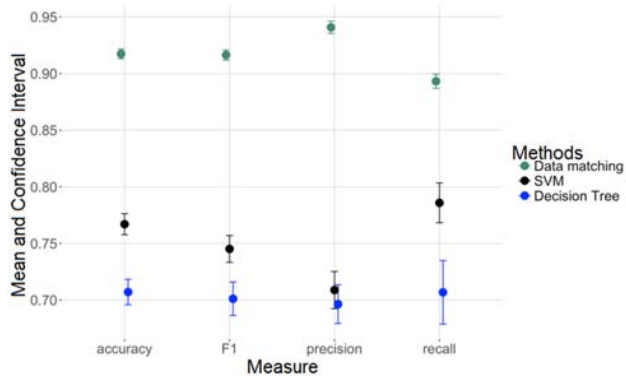

**Figure 2: Comparison of methods across measures of evaluation.** SVM and Decision Tree were the best performing algorithms for method 1 and 2, respectively.

## 2.3. PREPROCESSING: REMOVING NON-PERSON ACCOUNTS

Not all Twitter accounts belong to individuals and therefore should not be assigned a gender. To address this issue, we used labels available within the data provided by Kaggle that denote whether an account belongs to a person or brand to develop a prediction algorithm that removes non-persons. We used a set of profile features, including, friends count, followers count, mention of the terms "I" or "we" in the profile description, count of emoji in the profile description, length of profile description, provision of a non-social media URL, whether the user's name is found in historical US SSA data, and the average number of tweets issued per month to predict person or non-person status. We tested these features using an SVM classifier, a decision, tree, a random forest classifier, as well as simple and weighted ensembles of these three algorithms. We found that the random forest classifier performed best (accuracy=0.880, precision=0.904, recall=0.941, and F1=0.922). Applying this classifier to our foodborne data resulted in the removal of 2,684 non-person users.

## 3. RESULTS

We found that the most reliable and best performing algorithm for the supervised learning method using word and character n-grams (i.e. Method 2) was an SVM classifier with a linear kernel. For supervised learning and linguistic structure of names (i.e. Method 3), a decision tree performed best (see Table 1).

Also, as discussed in Section 2.2.4., the weighted ensemble classifier achieved a higher accuracy, precision, recall and F1 scores than the individual machine learning methods. Additionally, although the data matching method had a stronger performance than the ensemble approach, it only covered 68% of users. The ensemble method covered all users. Overall, the ensemble classifier was a more effective means of identifying users' gender than any individual method tested. These results illustrate that an ensemble approach using only the user's name can provide an accurate and reliable estimate of gender.

In our application to tweets on foodborne illness, we found that the West Coast – particularly, Southern California – has a high-volume of foodborne illness reports (see Figure 4). The majority of users within counties with high-volume tweet activity were female (see Figures 5-6). For instance, an estimated 57% of users who posted about foodborne illness from Los Angeles County – the county with the highest tweet volume in our data – were female. Indeed, we note that tweets within eight of the ten counties with the highest tweet volume in California – Los Angeles, Orange, San Bernardino, Riverside, Sacramento, Santa Clara, Alameda, and Fresno – were generated by 56% to 67% female Twitter users. Based on the 2010 US census, the proportion of male to female residents in these counties is comparable. See Figure 5 for a visual comparison of the gender distribution of the 2010 US census (a) and the gender distribution of the users tweeting (b).

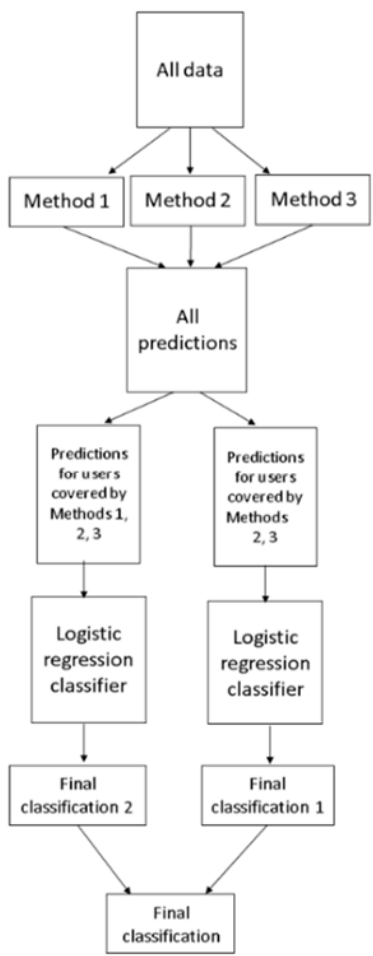

**Figure 3: Analysis framework**

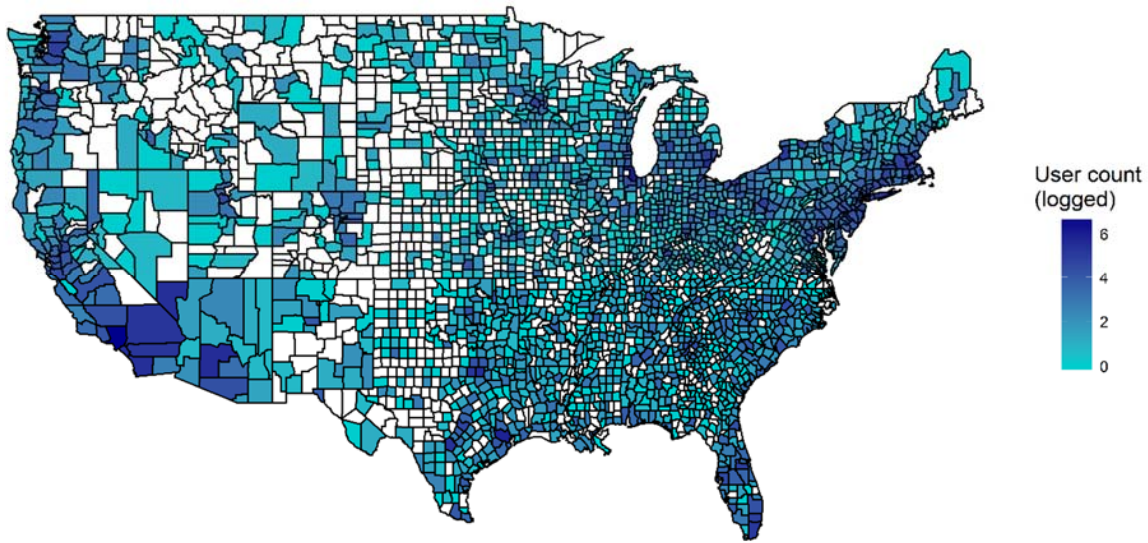

**Figure 4: Volume of users reporting foodborne illness**

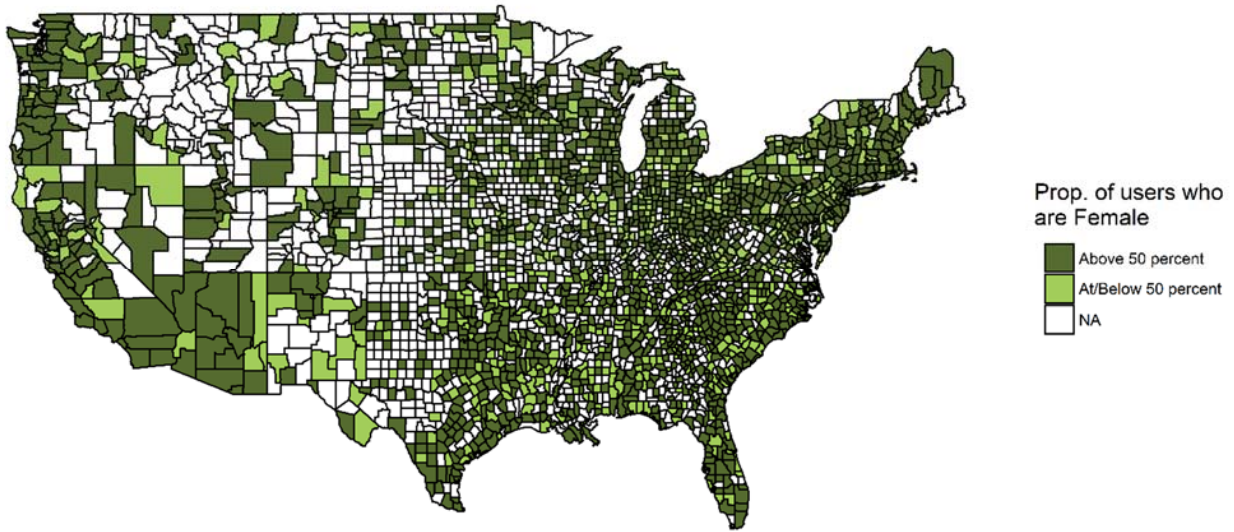

**Figure 5: Gender distribution of users reporting foodborne illness**

## 4.DISCUSSION

This study illustrates that simple user metadata can be used to build highly accurate and scalable demographic prediction models. We present an ensemble classifier that uses only user names to predict the gender of Twitter users with 82.8% accuracy and 85.2% recall.

We then illustrate that this approach can be used to quickly assess the gender distribution of a dataset containing 36,085 geotagged tweets posted by 33,208 users reporting foodborne illness in the United States.

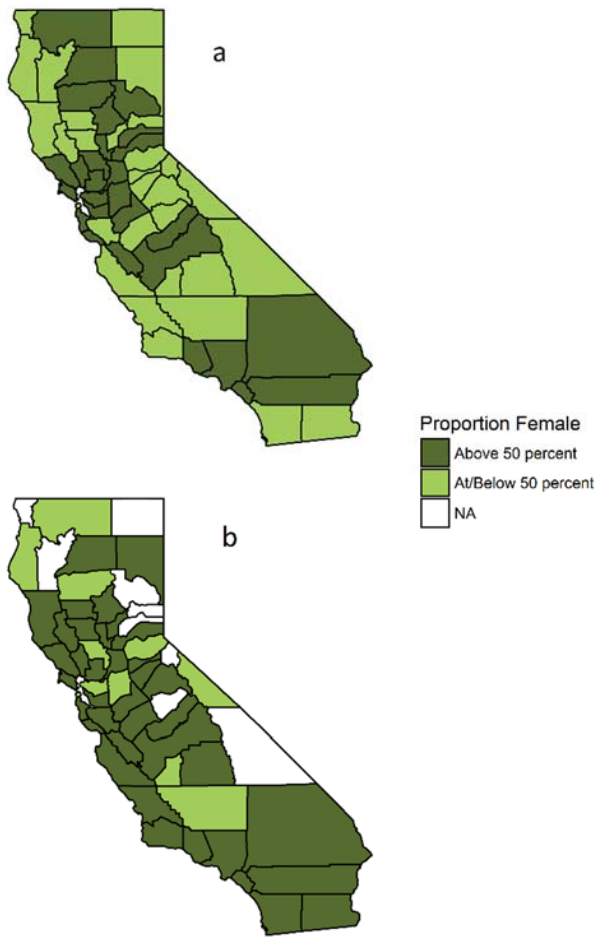

Figure 5: Gender distribution of counties (a), users tweeting within counties (b) in California

Knowing the demographic distribution of social media users is important for several reasons. First, research suggests that there may exist disparities in the demographic composition of individuals who use social media as a tool for reporting health status/behavior or discussing health related-issues [8, 11, 20, 24, 26, 28, 29]. The availability of demographic information would allow researchers and practitioners to adjust for these biases when using these data. Furthermore, researchers can also study gender differences in attitudes and responses to topics related to health and health care.

Research suggests that men and women use social media spaces in different ways. While men have caught up to women in regard to overall social media use [3], many sites are still more popular among female users. Furthermore, research suggests that male and female users engage with social media sites for different reasons. Women tend to use social media to sustain personal relationships and share personal updates, whereas men are more likely to seek out new connections and discuss abstract topics [3, 34]. These different usage patterns could render users more or less likely to be included in a keyword-based sample about health-related behaviors or attitudes.

Furthermore, the heavily female-generated data in our sample may be an artifact of gender-dependent patterns in the sharing of health information on Twitter. A survey we conducted of 1,649 Twitter users indicated that 38% of women and 25% of men tweet about their health.

### 4.1. DIRECTIONS FOR FUTURE RESEARCH
In future work, we plan to develop ensemble classifiers for predicting age and race of social media users. While these traits appear to be more difficult to predict than gender - for instance, studies predicting age tend to conflate numeric age with life stage [36] - there may be simpler approaches for predicting age that are yet to be explored. We also plan to develop open source tools that will allow researchers to upload a sample of Twitter data and explore the demographic composition of the sample.

### 4.2. PRIVACY AND ETHICAL CONCERNS
There are privacy and ethical concerns associated with the use of social media data for research as noted in several publications [9, 45, 53], and the development of tools for inferring user demographics. Since research subjects no longer participate in studies in a traditional sense [4], the treatment of human subjects is an important concern. It is important for us to be considerate of user privacy both on an individual and group level. Linking social media data to other online information or providing too many pieces of otherwise anonymous personal information increases the ability to de-identify users [41, 55]. Disclosure of group status – demographic or otherwise, also has the potential to facilitate de-anonymization, as well as invite biased treatment of a particular group given their online attitudes and behaviors [19, 56]. We seek to address these concerns by displaying only aggregated demographic distributions that do not link users to particular subtopics and sentiments, and by controlling access to our study data. Our aim is to allow researchers using social media for public health to control for demographic biases in their data, not to promote differential group treatment based on tweet content.

## 5. CONCLUSION
Studies have suggested that there exist demographic disparities in how individuals communicate about health on social media, so understanding the demographic composition of a social media sample could allow researchers to adjust for these biases. We address this challenge by developing a scalable, accurate and reliable approach to predicting gender that uses only users' names. Our ensemble methods can be used to quantify demographic disparities within data samples, which can improve our understanding of how different groups use social media so as not to amplify existing health disparities. Ongoing research is focused on developing similar approaches for the prediction of age and race.

## 6. ACKNOWLEGEMENTS
This work was funded by the Robert Wood Johnson Foundation (grant #73362).